\documentclass{PoS}
\usepackage{epsfig,graphicx,bm}
\usepackage{amssymb,amsmath}

\def\be{\begin{equation}}
\def\ee{\end{equation}}
\def\bea{\begin{eqnarray}}
\def\eea{\end{eqnarray}}
\def\ba{\begin{array}}
\def\ea{\end{array}}
\def\bmp{\begin{minipage}}
\def\emp{\end{minipage}}
\def\Li{\textrm{Li}}
\newcommand{\tr}{\textnormal{Tr}}

\newcommand{\delsl}{\partial\hspace{-2.1mm}/}
\def\half{{\textstyle{1\over 2}}}
\newcommand{\p}{{\bf p}}
\newcommand{\q}{{\bf q}}
\newcommand{\0}{{\bf 0}}

\title{Chiral phase transition in effective models of QCD}

\ShortTitle{Chiral phase transition in effective models of QCD}

\author{
\speaker{Zs. Sz{\'e}p}\thanks{The author thanks the organisers of the 
workshop for their kind invitation.}\\ 
Research Institute for Solid State Physics and Optics 
of the Hungarian Academy of Sciences, H-1525 Budapest, Hungary\\
E-mail: \email{szepzs@achilles.elte.hu}
}

\abstract{
Incorporating its most relevant global symmetry, chiral effective
models are aimed at investigating, in a simplified framework, important
aspects of the yet unsolved QCD. We review some recent results obtained in
these models on the restoration of chiral and axial symmetries, on the
boundary of the first order phase transition region in the
$m_\pi-m_K$--plane at vanishing baryonic chemical potential $\mu_B$, 
and on the $\mu_B-T$ phase diagram.}


\FullConference{29th Johns Hopkins Workshop on current problems in particle
theory: strong matter in the heavens\\
                 1-3 August\\
                 Budapest
}

\begin{document}

\allowdisplaybreaks{
\section{Introduction}

Although the issue of the deconfinement phase transition in strongly
interacting matter can be addressed with first principle calculation in
lattice field theory, it is of interest to see what can be obtained
regarding the chiral symmetry restoration aspect of the transition within
effective models sharing the same global symmetry as QCD.  For vanishing
chemical potential universality classes of the transitions of strongly
interacting matter were predicted by effective models for three corners of
the $m_{u,d}-m_s$--plane: the exact SU(3) chiral limit ($m_{u,d,s}=0$), the
two flavor chiral case ($m_{u,d}=0,$ $m_s=\infty$) and the pure gauge theory
($m_{u,d,s}=\infty$). As a consequence of the compatibility of the
conclusions there has to be also a tricritical point (TCP) on the
$m_{u,d}=0$ line \cite{Ukawa}. 
Apart from these corners and $m_{u,d}=0$ axis no universal arguments
exist and the order of the transition for different values of the
chemical potentials (baryonic, isospin) and quark masses, 
as well as the shape of the transition line between regions of crossover
and first order transitions are a matter of calculation.

It is expected that at vanishing baryonic chemical potential for low values
of the masses there is a line of second order phase transitions as a
boundary of the first order phase transition region. Increasing the value of
$\mu_B$ the second order line sweeps along a critical surface.  Theoretically
it is important to map the phase-diagram as a function of the parameters of
QCD, and also there can be experimentally accessible interesting points,
like the critical end point (CEP) occurring at finite $\mu_B$, for physical
values of the quark masses.

Apart from helping the mapping out of the QCD phase transition, the
effective models can give insights in the interplay between chiral and
$U_A(1)$ symmetry restoration, and how meson and baryon properties change
during the chiral transition. They can tell us what is the soft mode at the
CEP, and by predicting specific signatures they might help finding its
actual location \cite{signature}. They can also give information on the 
non-equilibrium dynamics near TCP or CEP \cite{Rajagopal,Patkos0}.

Effective models play also an important role whenever particle physics is
applied to cosmological or astrophysical studies. In cosmology, it is in
important to know what the cold dark matter is made of. The relic density of
one of the candidates, the weakly interacting massive particles (WIMPs),
depends on the variation of the number of relativistic degrees of freedom
and the energy density of the universe.  It was shown in \cite{Hindmarsh}
that if WIMPs freeze out in the range of the QCD transition temperature the
prediction of their relic concentration depends
on the QCD equation of state, which below $T_c$ is
taken from an effective model. In astrophysics the structure of compact
stars is actively investigated due to increasingly available new data that
indirectly provides information on the state of matter in their core. At
present effective models are the only tools for determining the equation of
state of strongly interacting matter at very low temperature and high
density.
 
In the present article a brief overview  of some of the effective models
will be given, and then some recent results on the issues enumerated in the
Abstract will be summarised.

\section{Effective chiral models\label{Sec:Eff}}

A common feature of the effective models is that they are constructed using
the global chiral symmetry of the massless QCD and its spontaneous symmetry
breaking pattern $SU_L(3)\times SU_R(3)\rightarrow SU_V(3)$. For equal
quark-masses the four-divergence of the vector current vanishes, while the
divergence of the axial vector current vanishes only in the massless case:
\bea
\partial_\mu\big[\bar{q}\gamma^\mu\lambda^aq\big]&=&
i\bar{q}[M,\lambda^a]q=
\bar q_i(m_i-m_j)\lambda^a_{ij}q_j, \\
\nonumber
\partial_\mu\big[ \bar{q}\gamma^\mu\gamma_5 \lambda^a q\big]&=&
i\bar{q}\{\lambda^a,M\}\gamma_5q=
\bar q_i(m_i+m_j)\lambda^a_{ij}\gamma_5 q_j,
\eea
where for three flavors $M=\mbox{diag}(m_u,m_d,m_s)$
and there is an order parameter (OP), the chiral condensate 
$\displaystyle
\langle 0|[Q^a_A, \bar q \gamma_5 \lambda^b q] |0\rangle=
-{1\over 2} \langle 0|\bar q \{\lambda^a,\lambda^b\} q |0\rangle =
-{2\over 3} \delta_{ab} \langle 0|\bar q q |0\rangle$,
which when vanishes indicates the restoration of chiral symmetry.

The Lagrangian contains explicit symmetry breaking terms of two kinds. One is
the quark mass term, or in the linear sigma model the term containing the
external sources, which gives mass to the pseudo-Goldstone bosons. The other
type is the instanton motivated 't Hooft determinant which breaks the
$U_A(1)$ symmetry.

The effective models are built upon mesons and/or baryons with the lowest
mass and are expected to be most relevant in case of a crossover transition
when the mesons are well defined degrees of freedom even above the
pseudo-critical temperature or in case of a second order phase transition
where the change of the order parameter is directly accessible.

The effective models differ essentially from QCD in that they miss the
mechanism for confinement and incorporate only the degrees of freedom
thought to be relevant for the chiral transition. Because of this latter
feature it is very important to have a faithful, fine-tuned parametrisation
of any effective model. I mention, however, that recently there have been
studies \cite{Sannino}, \cite{Fukushima_Polyakov}
that included confinement in chiral models through an effective
potential constructed with its (approximate) order parameter, the Polyakov 
loop, in an
attempt to give account of the lattice result which shows that both chiral
and deconfinement transitions happen at about the same
temperature \cite{Karsch_LN}.

\noindent {\bf The chiral perturbation theory--ChPT}
                             
Chiral perturbation theory \cite{ChPT1} describes exactly the low energy
dynamics of the Goldstone particles collected in a hermitian traceless 
matrix $\Phi(x)$:
\be
\Phi (x) =
\begin{pmatrix}
{\pi^0/\sqrt 2} + {\eta_8/\sqrt 6} & \pi^+ & K^+ \cr
\pi^- & - {\pi^0/\sqrt 2} +  {\eta_8/\sqrt 6} & K^0 \cr
K^- & \bar{K}^0 & - {2\eta_8/\sqrt 6}
\end{pmatrix}.
\ee
The most general Lagrangian invariant under $SU_L(3)\times SU_R(3)$ 
chiral symmetry is written down in terms of the exponential of $\Phi$, 
$U(\Phi)=e^{i\sqrt{2}\Phi/f}$
and appears as a systematic expansion in momenta and quark masses:
\be
{\cal L}^{SU(3)}_{\mathrm{ChPT}}(U) = \sum_n {\cal L}_{2n}
= {f^2\over 4} \mbox{Tr}(\partial_\mu U^\dagger \partial^\mu U )
+\frac{f^2 B_0}{2}\mbox{Tr}(MU^\dagger+UM^\dagger) +{\mathcal O} (p^4).
\label{Eq:L_ChPT}
\ee
In the general $U_L(3)\times U_R(3)$ case \cite{ChPT2} this is more 
complicated because each term of (\ref{Eq:L_ChPT}) can be
multiplied by some function of $\eta_0=-i f \log \mbox{det}
U/\sqrt{6}$ (in this case $U=e^{i(\sqrt{2}\Phi+\eta_0\lambda^0)/f}$). 
What is important for us from the point of view of the parametrisation 
of the linear sigma model in the $m_\pi-m_K$--plane is that ChPT determines 
the $m_\pi$ and $m_K$ dependence of the 
decay constants for pions ($f_\pi$) and the kaons ($f_K$) as well as of 
$M_\eta^2=m_\eta^2+m_{\eta'}^2$:
\begin{eqnarray}
\label{decayconst_pi}
f_\pi&=&f\left[1-\frac{1}{f^2}(2\mu_\pi+\mu_K-4m_\pi^2(L_4+L_5)-
8m_K^2L_4)\right],
\\
\label{decayconst_K}
f_K&=&f\left[1-\frac{1}{f^2}\left(\frac{3}{4}(\mu_\pi+\mu_\eta+2\mu_K)-
4m_\pi^2L_4-4m_K^2(L_5+2L_4)\right)\right],
\\\nonumber
{M_\eta^2}&=&2m_K^2-3v_0^{(2)}+2(2m_K^2+m_\pi^2)
(3v_2^{(2)} - v_3^{(1)}) 
+\frac{1}{f^2}\Big[8v_0^{(2)}
(2m_K^2+m_\pi^2)(L_5+3L_4)\\ \nonumber
&&
+m_\pi^2(\mu_\eta-3\mu_\pi)-4m_K^2\mu_\eta
+\frac{16}{3}(6L_8-3L_5+8L_7)(m_\pi^2-m_K^2)^2\\&&
+\frac{32}{3}L_6(m_\pi^4-2m_K^4+m_K^2m_\pi^2)+
\frac{16}{3}L_7(m_\pi^2+2m_K^2)^2 \Big].
\label{Meta-extrap}
\end{eqnarray}
where
$\displaystyle \mu_{PS}:=\frac{m_{PS}^2}{32\pi^2}\ln\frac{m_{PS}^2}{M_0^2}$
are the so--called chiral logarithms, determined at an appropriate
scale (e.g. $M_0=4\pi f$) 
by the masses of the members of the pseudoscalar octet. $f$ is the pion decay
constant in the chiral limit. The low energy constants $L_i$ are determined 
by the values of the masses and decay constants taken in the physical point.

\noindent {\bf The Nambu--Jona-Lasinio (NJL) model}

The Lagrangian of the chiral $SU(3)$ symmetric NJL model is written 
in terms of the current quarks with mass matrix 
$\hat m=\textrm{diag}(m_u,m_d,m_s)$ 
\be
{\mathcal{L}}_{\mathrm{NJL}}^{SU3}=
\bar{q}(i\delsl-\hat m)q+
\frac{g_S}{2}\sum_{a=0}^8[(\bar{q}\lambda^a q)^2+
(\bar qi\gamma_5\lambda^{a}q)^2]+
g_{D}\big\{\mbox{det}[\bar q (1+\gamma_5) q]+
\mbox{det}[\bar{q}(1-\gamma_5)q]\big\}.
\ee
The determinant term breaks the $U_A(1)$ symmetry while the second term is 
invariant under chiral U(3) group transformations. 
For two flavors the determinant term can be rewritten as
$$2\big\{
\mbox{det}[\bar q (1+\gamma_5) q]+
\mbox{det}[\bar{q}(1-\gamma_5)q]\big\}=
\big\{ ({\bar q}q)^2- ({\bar q} \vec\tau q)^2
- ({\bar q} i\gamma_5 q)^2
+ ({\bar q} i\gamma_5\vec\tau q)^2 \big\},$$
and by taking equal couplings $g_D=g_S\equiv G$
one arrives at the original form of the NJL 
Lagrangian \cite{NJL_orig}:
\be
{\cal L}_{\mathrm{NJL}}^{SU2}=\bar q(i\delsl-\hat m) q +
G\big\{(\bar q q)^2 + (\bar q i\gamma_5\vec\tau q)^2 \big\}.
\ee 
This is invariant under axial transformations 
$q\rightarrow\big(1-\half i\vec\alpha\cdot\vec\tau\gamma_5\big)q$
since $\bar q q\rightarrow \bar q q-
\vec\alpha\cdot\bar q i\vec\tau \gamma_5 q$ and
$\bar q i\vec\tau\gamma_5 q\rightarrow\bar q i\vec\tau\gamma_5 q+
\vec\alpha\bar q q$.
In the mean-field approximation the square of the quark bilinear is
linearised with the replacement
$(\bar q q)^2\rightarrow 2\langle \bar q q\rangle (\bar q q)-
\langle \bar q q\rangle^2$
and the vacuum expectation value of the quark bilinear, the order
parameter, determines
the constituent quark mass $M$ for which the self-consistent
gap-equation reads:
\be
M=m_{ud}-2G\langle\bar q q\rangle=
m_{ud}+2N_c N_f  G\int_{|\p|<\Lambda}\frac{M}{\sqrt{\p^2+M^2}}
\left[1-n_q(T,\mu)-n_{\bar q}(T,\mu)\right].
\ee                                                                          
The natural, mesonic degrees of freedom $\pi, \sigma$ are obtained through the
procedure of bosonisation. Monitoring the expectation value 
$\langle\bar q q\rangle$, one can study the restoration of chiral symmetry. 
For a review of recent developments in the NJL model see \cite{NJL} and 
references therein. 

\noindent {\bf The linear sigma model (L$\sigma$M)}
                       
The Lagrangian of the linear sigma model with symmetry breaking terms
is constructed in terms of a $N_f\times N_f$ complex matrix 
$M=T_a (\sigma_a +i \pi_a)$ which contains scalar ($\sigma_a$) 
as well as pseudo-scalar ($\pi_a$) fields \cite{LSM}. For $N_f\le 4$
the Lagrangian reads:
\be
{\mathcal{L}}_M={\rm tr}(\partial_\mu M^{\dag} \partial^\mu M+\mu_0^2
M^{\dag} M)-f_1{\rm tr}^2(M^{\dag} M)-f_2 {\rm tr}(M^{\dag}M)^2
-g\big(\det M+\det M^{\dag}\big)
+{\rm tr} H (M+M^{\dag}).
\label{Lagrangian}
\ee
For $N_f=3$ the generators $T_a$ are proportional to the Gell-Mann matrices. 
There are two independent quartic couplings $f_1$ and $f_2$. 
The explicit symmetry breaking term contains the external fields $h_a$
($H=T_a h_a$) which gives mass to the pseudo-Goldstone bosons. The role of the 
determinant term is to break the chiral $U(N_f)$ symmetry down to $SU(N_f)$.

In the broken symmetry phase the scalars associated with the diagonal
generators display vacuum expectation values. Considering only the case of
degenerate u, d quarks one has for two, three, and four flavors one, two, or
three condensates, called non-strange, strange and charmed condensates,
respectively.

It is important to stress that the determinant term enters with a relative
opposite sign into the expressions of the tree-level masses of scalars and
pseudo-scalars:
\bea
\nonumber
\left[m_S^2(\bar\sigma)\right]_{ab}&=&\mu_0^2 \delta_{ab} -
6[\delta_{N_f,2}\ {\cal G}_{ab}+\delta_{N_f,3}\ {\cal G}_{abc} \bar\sigma_c ] +
4[{\cal F}_{abcd}+ \delta_{N_f,4}\ {\cal G}_{abcd}]
\bar\sigma_c \bar\sigma_d, \\
\left[m_P^2(\bar\sigma)\right]_{ab} &=&\mu_0^2 \delta_{ab}+
6[ \delta_{N_f,2}\ {\cal G}_{ab} +\delta_{N_f,3}\ {\cal G}_{abc} \bar\sigma_c]+
4[{\cal H}_{abcd}\,-\,\delta_{N_f,4}\ {\cal G}_{abcd}]
\bar\sigma_c \bar\sigma_d.
\label{Eq:masses_tree}
\eea
The various tensors appearing in (\ref{Eq:masses_tree}) are explicitly 
given in \cite{Roder}.
The parameters of the model are determined using the zero temperature
mass spectra and decay constants. For the case $N_f=3$ this will be
explained in Section \ref{sec:boundary}.

\noindent {\bf O(N) model}

For two flavors the 8 d.o.f. of the  complex matrix $M$ introduced previously 
form a reducible representation of $SU_L(2)\times SU_R(2)$. The irreducible 
representations are given by two O(4) multiplets $(\sigma,\pi)$
and $(a_0,\eta)$, whose masses are split by the determinant term:
$m_\eta^2-m_\pi^2=2g,$ $m_{a_0}^2-m_\sigma^2=2g-2f_1\bar\sigma_0$.
Assuming that the $(a_0,\eta)$ multiplet is heavier 
(formally $g\to \infty$) one can neglect it, 
(integrate it out) and one arrives at the O(4) model.
An equivalent formulation is to use instead of general complex matrices
unitary matrices with positive determinant 
$M=\sigma{\bf 1}+i\vec{\pi}\cdot\vec{\tau}$ written with the help of  
sigma and pion fields \cite{Wilczek}. If one increases the number of pions 
and rescales the couplings in such a way that the energy density is 
proportional to the number of  d.o.f. per site  ($\sim N^2$) and the 
masses stay finite ($\sim N^0$),
one obtains a form of the Lagrangian suitable for a large N
approximation (see the meson part in the Lagrangian (\ref{Eq:L_CQM}), below).

\noindent {\bf The chiral constituent quark model}

This model extends the L$\sigma$M by including some effective fermionic 
degrees of freedom, the constituent quarks. For a large N treatment 
$SU_L(2)\times SU_R(2)$ L$\sigma$M model 
is extended to $N_f$ flavors, that is the number of pions is increased and 
the appearance of an other quartic coupling is disregarded. 
The constituent quark mass is generated by the scalar condensate.
In the broken symmetry phase characterised by the vacuum expectation value 
$\Phi$, the Lagrangian is parameterised in a way to be used in a large N 
approximation. Fermions couple with a Yukawa coupling to 
the mesons:
\bea
{\mathcal{L}}_{\mathrm{CQM}}&=&-\left[\frac{\lambda}{24}\Phi^4+\frac{1}{2}m^2\Phi^2\right]N-
\left[\frac{\lambda}{6}\Phi^3+m^2\Phi+h\right]\sigma\sqrt{N}
+\frac{1}{2}\bigl[(\partial^\mu\sigma)^2+(\partial^\mu\pi^a)^2\bigr]
\nonumber\\
&-&\frac{1}{2}\bigl[m^2+\frac{\lambda}{2}\Phi^2\bigr]\sigma^2
-\frac{1}{2}\bigl[m^2+\frac{\lambda}{6}\Phi^2\bigr]\pi^a\pi^a
-\frac{\lambda}{6\sqrt{N}}
\Phi\bigl[\sigma^3+\sigma\pi^a\pi^a\bigr]-
\frac{\lambda}{24N}\bigl[\sigma^2+\pi^a\pi^a\bigr]^2
\nonumber\\
&+&
\bar \psi(x)\left[i\partial^\mu\gamma_\mu-m_q-
\frac{g}{\sqrt{N}}\left(
\sigma (x)+i\sqrt{2N_f}\gamma_5T^a\pi^a(x)\right)\right]\psi(x),
\label{Eq:L_CQM}
\eea
where $m_q=g\Phi$ and $\Phi(0)=f_\pi/2$. This parametrisation was obtained
by requiring that the fermion mass 
stays finite as $N\to \infty$. We can see that due to the large N 
treatment the fermion coupling to the pions is enhanced over the sigma. 
The fermion contribution is of order $1/\sqrt{N}$ and thus it precedes 
in an $1/N$ expansion the NLO meson contribution, which is of order $1/N$.

\section{Solving the effective models, thermodynamics}

Since the perturbation theory fails to account for non-trivial
features of phase transitions, resummation of perturbative series is needed. 
Such method is the optimised perturbation theory
(OPT) proposed in \cite{Chiku_Hatsuda}. It consists of introducing into the
Lagrangian a temperature dependent mass with the change $-\mu^2_0 \to M^2(T)$ 
and treating the difference between this and the original one as a 
finite counter-term $\Delta m^2=M^2(T)+\mu_0^2$:
\be
{\mathcal{L}}_{\mathrm{mass}}=-\frac{1}{2}M^2(T)\tr\ \Phi^\dag \Phi+
\frac{1}{2}\Delta M\tr\ \Phi^\dag \Phi.
\ee
The introduced mass-term is fixed through the Principle of Minimal Sensitivity
by requiring that the one loop pion-mass defined at zero external momentum, 
calculated using tree-level masses in the propagators appearing in the
self-energy, equals the tree level mass for the pions:
\be
M_{\pi}^2(T)=\left. iG^{-1}({\bf p=0},T)\right|_\textnormal{1-loop}=
\left.-m_\pi^2(T) \right|_\textnormal{tree} -\Delta m^2-
{\displaystyle \sum_{\displaystyle{k\,l}}}
\raisebox{-0.4cm}{\includegraphics[width=4cm]{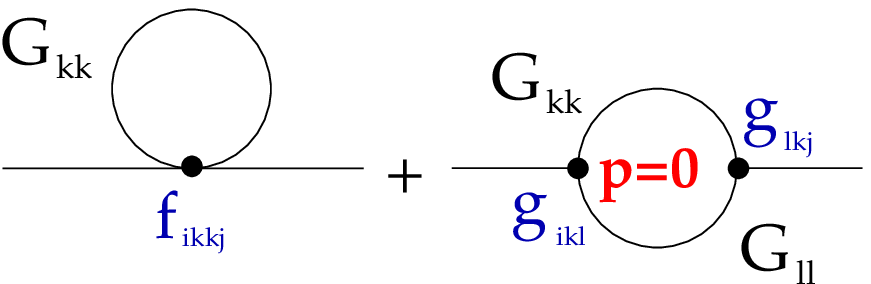}}
\left.
\overset{\displaystyle!}= m_\pi^2(T) \right|_\textnormal{tree}
\label{mass_def}
\ee
This results in a self-consistent gap equation for the pions, since all 
the other tree level masses can be expressed with the pion mass and the 
vacuum expectation values:
\be
m_\pi^2(T)=-\mu_0^2+2(2f_1+f_2)x^2+4f_1y^2+2gy+\sum^{\alpha=
\sigma,\,\pi}_{i=\pi,\,K,\,\eta,\eta^{'}} c^\pi_{\alpha_i} 
\textnormal{I}_{\textrm{tp}}(m_{\alpha_i}(T)) \, .
\label{pigap}
\ee
The summation is understood over the members of both the scalar and
pseudoscalar multiplets.

In oder to determine the order of the phase transition the gap-equation 
(\ref{pigap}) has to be solved simultaneously 
with the equations of state which read as:
\begin{eqnarray}
\label{eq:EoS_x}
&-\epsilon_x-\mu_0^2 x+2gxy+4f_1xy^2+2(2f_1+f_2)x^3+
\sum^{\alpha=\sigma ,\pi}_{i=\pi,\,K,\,\eta,\eta^{'}}  J_i t^x_{\alpha_i} 
\textnormal{I}_{\textrm{tp}}(m_{\alpha_i}(T))&=0 \label{xeq}\,,\\
\label{eq:EoS_y}
&-\epsilon_y-\mu_0^2 y+gx^2+4f_1x^2y+4(f_1+f_2)y^3+
\sum^{\alpha=\sigma ,\pi}_{i=\pi,\,K,\,\eta,\eta^{'}}  J_i t^y_{\alpha_i} 
\textnormal{I}_{\textrm{tp}}(m_{\alpha_i}(T))&=0\,, \label{egy2}
\end{eqnarray}
where the weights\  $t^x_{\alpha_i}$ and $t^y_{\alpha_i}$ of the tadpoles
and the isospin multiplicity factors $J_i$ are listed in Appendix C of 
\cite{Patkos3}. The renormalization was discussed in
\cite{Chiku_Hatsuda} (see also \cite{Toni}).
With the definition of the mass given in (\ref{mass_def}) only tadpole type
integrals appear, and the only complication is due to the mixing sector
where the eigenstates are the $\eta$ and $\eta'$, and $\sigma$ and $f_0$
respectively.  Because OPT does not modify the form of the relations upon
which the Goldstone theorem relies, this method preserve the Goldstone
theorem for the pions. However, as explained in \cite{Patkos3} resumming
only one parameter in this model with two order parameters has the
consequence that the Goldstone theorem for kaons is not satisfied at
finite T.

%\medskip

Another (approximate) solution of the linear sigma model was given using the
Hartree approximation within the CJT formalism \cite{CJT}, see also
\cite{Petropoulos}. Using Euclidean metric, we illustrate this method for 
the case of the O(N) model with an explicit symmetry breaking external
field, in the large N approximation. For the chiral $SU(N)$ symmetry there
is an additional complication due to the mixing sectors. The effective
potential parameterised with the vacuum expectation value $\Phi$ and the
dressed pion propagator $G_\pi$, includes all the diagrams that do not
become disconnected upon cutting two lines (in the Hartree approximation
only the pion ``double scoop'' diagram arises)
\bea
\nonumber
V[\Phi,G_\pi]&=&N\left[\frac{m^2}{2}\Phi^2+\frac{\lambda}{24}\Phi^4
-h\Phi\right]
+\frac{N-1}{2}\left[\int_\beta\ln G_\pi^{-1}(\Phi;k) 
+\int_\beta\Big[{\cal D_\pi}^{-1}(\Phi;k) G_\pi(\Phi;k)-1\Big]\right]\\
&&
+\frac{\lambda (N^2-1)}{24 N} \left[\int_\beta G_\pi(\Phi;k)
\right]^2.
\eea
The vacuum expectation value and the dressed propagator are determined from the
two stationarity conditions 
$\displaystyle \frac{\delta V[\Phi,G_\pi]}{\delta G_\pi}=0, 
\frac{\delta V[\Phi,G_\pi]}{\delta \Phi}=0$.
The first one gives a Dyson-Schwinger equation
\be
G_\pi^{-1}(\Phi;k)={\cal D_\pi}^{-1}(\Phi;k)+\frac{\lambda}{6}
\int_\beta G_\pi(\Phi;k)=
k^2+m^2+\frac{\lambda}{6}\Phi^2+\frac{\lambda}{6}
\int_\beta G_\pi(\Phi;k),
\label{gap-CJT}
\ee
and since the self energy is momentum independent, 
with the parametrisation $\displaystyle G_\pi(\Phi;k)=\frac{1}{k^2+M_\pi^2}$
 one arrives at a gap equation for the resummed pion mass:
$\displaystyle 
M_\pi^2=m^2+\frac{\lambda}{6}\Phi^2+\frac{\lambda}{6}
\int_\beta\frac{1}{k^2+M^2_\pi}.$

The second condition gives the equation of state:
\be
\displaystyle
\left[m^2+\frac{\lambda}{6}\Phi^2+\frac{\lambda}{6}\int_\beta\frac{1}{k^2+M^2}
\right]\Phi=h.
\label{EoS-CJT}
\ee 
We see from the gap equation (\ref{gap-CJT}) and the equation of state 
(\ref{EoS-CJT}) that in the large N approximation $M_\pi^2=h/\Phi$, 
that is Goldstone's theorem is satisfied. This is not the case in general
because it is known that without the large N approximation Hartree
approximation has two problems: it gives first order (instead of second
order) phase transition in the chiral limit and Goldstone's theorem is not
fulfilled.

%\medskip

The last method for solving an effective model which we discuss now is the
large N approximation applied here to the chiral constituent quark model of
Sec.~\ref{Sec:Eff}.  We have solved this model in the leading order of the
large N approximation by taking the fermions into account perturbatively at
one loop order \cite{Patkos2}, although there is an infinite series of 
diagrams containing fermions and giving contribution at order $1/\sqrt{N}$
of the large N expansion.

Imposing a gap equation for the pions which performs the resummation of
daisy--type diagrams one obtains the temperature dependence of the order
parameter by simultaneously solving the gap equation and the equation of
state:
\bea
\nonumber
\left.
\begin{array}{l}
\displaystyle
\textnormal{gap equation:} \hspace*{0.6cm}
M^2=-iG_\pi^{-1}(p=0)=
m^2+\frac{\lambda}{6}\Phi^2+\frac{\lambda}{6}T_B(M)-4\frac{g^2N_c}{\sqrt{N}}
T_F(m_q)
\\
\displaystyle
\textnormal{EoS:}\hspace*{2.0cm}
V_{\mathrm{eff}}'(\Phi)=\Phi\left[m^2+\frac{\lambda}{6}\left(\Phi^2+T_B(M)\right)
-\frac{4g^2N_c}{\sqrt{N}}T_F(m_q)\right]-h=0
\end{array}\right\}\Rightarrow \hfill
\raisebox{0.75cm}{\textnormal{\parbox[t]{1.8cm}{Goldstone\\
    theorem\\ $\displaystyle M^2=\frac{h}{\Phi}$. }}}
\hfil
\eea
$T_{B,F}$ are the temperature-dependent 
pion and fermion tadpole integrals, respectively 
($m_q=g\Phi$). 

The form of the $\sigma$ propagator can be inferred from the
consistency condition
$-iG_\sigma^{-1}(p=0)=d^2 V_{\mathrm{eff}}(\Phi)/d\Phi^2$:
\be
\displaystyle
G_\sigma^{-1}(p)=p^2-\frac{h}{\Phi}-
2\Phi^2\frac{\frac{\lambda}{6}-\frac{4g^4N_c}{\sqrt{N}} I_F(p,m_q)}
{1-\frac{\lambda}{6} I_B(p,M)},
\ee
where $I_{B,F}$ are the temperature-dependent
pion and fermion one-loop ``fish'' integrals.
Analytically continuing the sigma propagator to the second Riemann sheet
the couplings were determined by requiring that the ratio of the real and
imaginary parts of the sigma pole on this sheet is close to
phenomenologically expected ratio of the mass to width of the sigma 
particle $m_\sigma/\Gamma_\sigma\approx 1$, while keeping the mass of 
the sigma in the range $m_\sigma\in (400,800)$ MeV.
The Yukawa coupling g is fixed by the constituent quark mass (one
third of the nucleon mass) and $f_\pi$: $g=m_q/\Phi(0)=6.72$.
Consistency of the $T=0$ $\sigma$-pole with 
phenomenological values of $m_\sigma$ and 
$\Gamma_\sigma$ requires $\lambda\simeq400$.

Once the couplings are fixed one returns to the solution at finite
temperature/density.

\section{Restoration of chiral and axial symmetry}

We start the discussion with the effects of the restoration of chiral symmetry
as felt by the sigma excitation.

In an effective theory the sigma particle is the quantum fluctuation of the
amplitude of the chiral order parameter, hence any change in the ground
state of the system is reflected upon its properties. Sigma couples strongly
to the pions, measurable through the $\sigma \to 2\pi$ decay.  Because of this,
at $T=0$ sigma shows up as a broad resonance in the scalar--isoscalar
channel. Hatsuda and Kunihiro conjectured \cite{Hatsuda1} that since the
sigma mass decreases during chiral symmetry restoration, the available phase
space for $\sigma \to 2\pi$ squeezes, therefore one has a chance to
see the sigma as a sharp resonance in matter. Observing the changes in the
sigma one can infer features of the chiral symmetry restoration.
\begin{figure}[htpb]
\centering{
\includegraphics[width=0.6\textwidth]{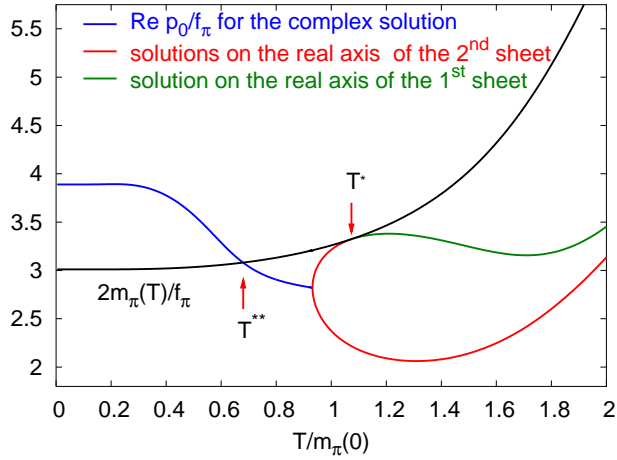}
}
\caption{The temperature dependence if the real part of the sigma pole.}
\end{figure}

Following with the increasing temperature the trajectory of the complex 
sigma pole on the second Riemann sheet obtained in the O(N) model, 
with the help of the LO of the large N approximation, 
one can see that the real part of the sigma goes below the value of the two 
pion threshold at $T^{**}\approx 0.69 m_\pi(0)=96.6$ MeV, which means the 
suppression of the $\sigma \to 2\pi$ decay channel \cite{Patkos1}. 
This happens at a lower temperature than 
$T^{*}\approx 1.07 m_\pi(0)$, where $\sigma$ is converted into a
stable particle. So, for $T<T^{**}$ sigma is indeed a 
sharper resonance compared with the vacuum case.
This scenario differs from the one in \cite{Hidaka} 
where the pole that goes through the threshold is not connected with 
the complex sigma pole at $T=0$. By assuming certain analytic
properties of the propagator, we have shown that there are universal 
features of the spectral function when a pole corresponding to a
particle at high temperature approaches in the complex energy plane
with the variation of the temperature  the threshold position of its
two-body decay \cite{Patkos1c}.

It is interesting to note, that the value of the temperature where 
$m_\sigma=2m_\pi$  is quite close to what was found within the
composite operator formalism applied to the QCD: 
$T^{**}=0.95 T_c\simeq 98$ MeV \cite{Barducci_T}.
Another interesting point is that within the large N formalism the mass of 
the sigma appears to be rather low (at NLO it is about 350 MeV,
\cite{Andersen}).

%\medskip

The restoration of the chiral and axial symmetries was studied recently 
by many authors, using both the linear sigma model
\cite{Lenaghan1,Lenaghan2},\cite{Roder} and NJL \cite{costaUA1big}.
The method of these studies was to identify the chiral partners 
and observe the convergence of their masses. The chiral partners are 
$(\pi,\sigma)$, $(\eta,a_0)$ in the case of chiral SU(2) symmetry 
and $(\pi,a_0)$, $(K,\kappa)$, $(\eta,\sigma)$ in the case 
of chiral SU(3) symmetry. 
The results of these investigations can be summarised as follows.

The scalar masses decrease at first but then start to increase during
the transition while the pseudoscalar masses stay constant for a long while
then increase monotonically with the temperature during the transition. The 
masses grow more rapidly when $N_f$ increases, and there is a 
tendency of the chiral partners towards degeneracy at some $T>T_c$.
In the absence of $U_A(1)$ anomaly $T_c$ decreases, and the degeneracy 
is more rapidly reached. The melting of the strange condensate is much
slower compared with the non-strange one.

For the effect of the $U(1)$ anomaly it was found that in the chiral limit
with(out) $U_A(1)$ anomaly the transition temperature increases (decreases)
with $N_f$ in contrast to (in agreement with) the QCD lattice results which
according to \cite{Karsch_LN} give for $N_f=2 (3)$ $T_c\simeq173 (154)$ MeV.
This signals that the anomaly is rapidly decreasing near the transition
temperature. Indeed, this decrease can be observed, since the finite
temperature lattice result on topological susceptibility \cite{Alles}
can be converted into the temperature-dependence of the strength of the
determinant  term, $g_D(T)$ by fitting the lattice result with the explicit 
formula of the susceptibility calculated in NJL \cite{Fukushima}. 
In this case the determinant term decreases not only due to the very slow 
melting of the  strange condensate, but also due to the  decrease of 
$g_D(T)$. Hence $U_A(1)$ symmetry is restored and the degeneracy of 
the SU(3) chiral partners is completed. Without the restoration of the 
$U_A(1)$ symmetry the degeneracy among parity partners
arises only at very high temperature at which 
the determinant term would eventually vanish due to the vanishing of 
the strange condensate. (We have to recall that the determinant enters 
with opposite sign into the expressions of the scalar and pseudoscalar masses.)

\section{$\mu_B-T$ phase diagram for $N_f=2$}

The left hand side figure of Fig.~\ref{Fig:mu-T} shows a generic 
phase diagram typical for those obtained with effective models. 
In particular this figure was obtained for the chiral limit, solved in 
the LO of the large N approximation, with the fermions taken into account
perturbatively at one loop order \cite{Patkos2}. The emerging diagram is
qualitatively correct,
but when compared with lattice results, one observes that the transition 
temperature is too low (the lattice result of \cite{Karsch_Tc} is 
$T_c(\mu_B=0)=173(8)$MeV) and the present estimate for
TCP is presumably located at a too high chemical potential.
\begin{figure}[tpb]
\hspace*{-0.2cm}
%\vspace*{-0.8cm}
\includegraphics[width=8cm]{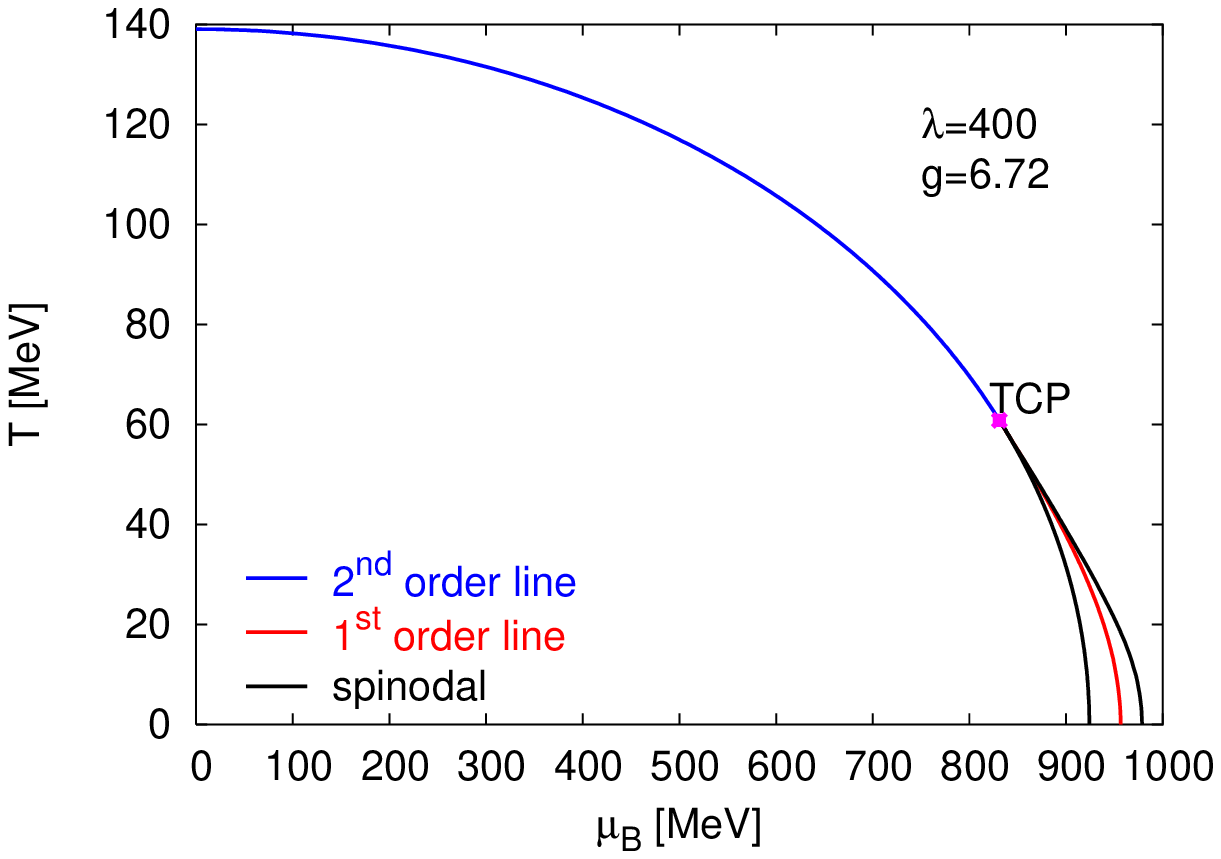}
\raisebox{0.07cm}{\includegraphics[width=7cm]{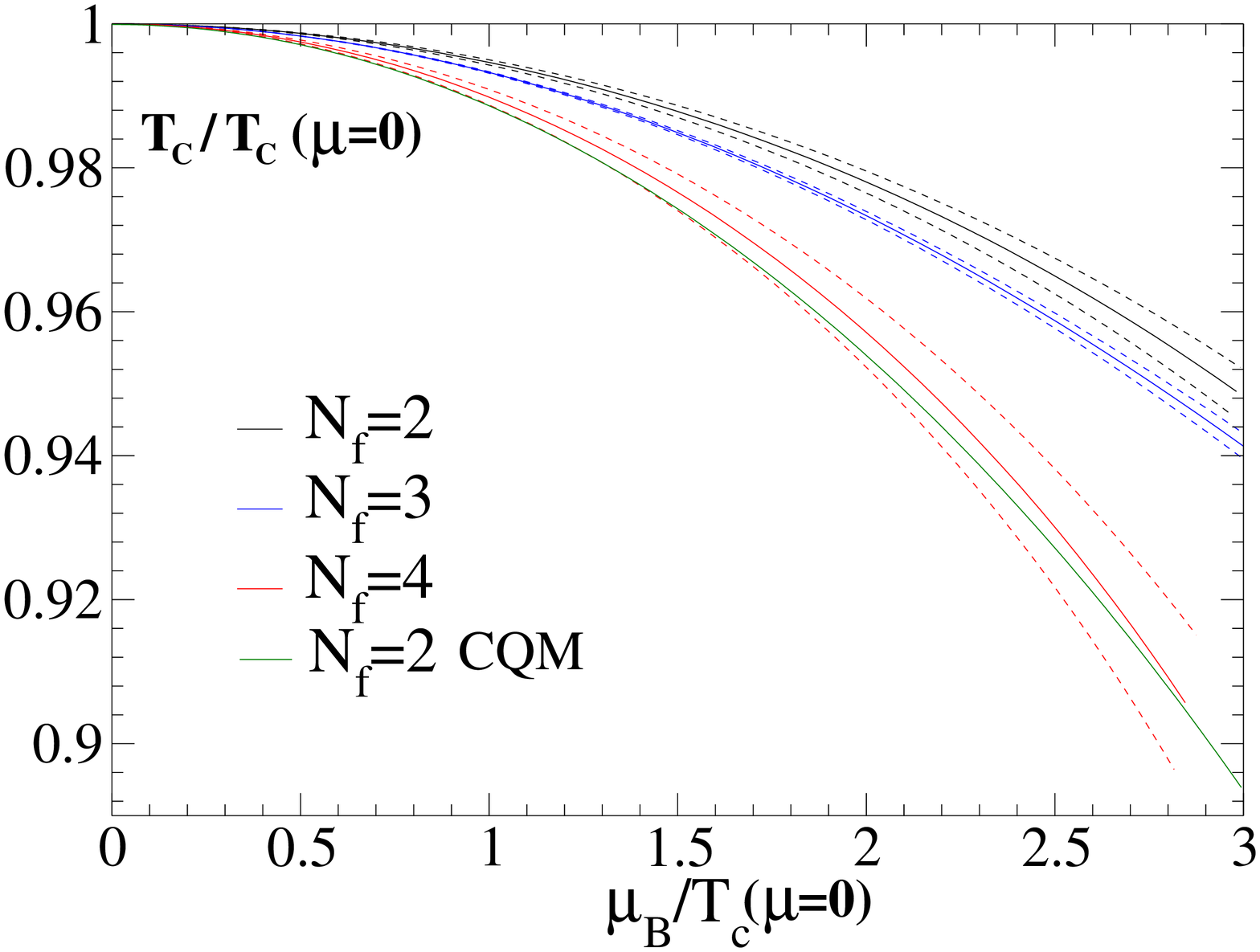}}
\caption{L.h.s.: The $\mu_B-T$ phase diagram of the constituent quark model
in the chiral limit.
R.h.s: Comparison of the CQM result from \cite{Patkos2}
with lattice results 
(from top to bottom): $N_f=2$ \cite{deForcrand1}, $N_f=3$
\cite{deForcrand2}, $N_f=4$ \cite{d'Elia} (the original figure was
taken from \cite{deForcrand2}).}
\label{Fig:mu-T}
\end{figure}
The transition line, which is a parabola, and the location of TCP 
can be obtained analytically in terms of polylogarithms:
\bea
\nonumber
\left.
\begin{array}{l}
{ 2^\textnormal{nd}\,
\textnormal{order+spinodal line}}\qquad
\displaystyle
m^2+\left(\frac{\lambda}{6}+g^2N_c\right)\frac{T^2_c}{12}+
\frac{g^2\mu^2_q}{2\pi^2 N_f}N_c=0,
\vspace*{0.2cm}
\\ \nonumber
{ 2^\textnormal{nd} \, \textnormal{order line ends
when}}\qquad
\displaystyle
\frac{\lambda}{6}+\frac{g^4N_c}{4\pi^2}\left[
\frac{\partial}{\partial n}
\Big(\Li_n\left(\frac{1}{z}\right)+\Li_n\left(z\right)\Big)\Big|_{n=0}-
\ln\frac{cT_c}{M_{0}}
\right]=0
\end{array}\right\}\Rightarrow
\parbox[t]{1.2cm}{$T_{TCP}$\\ $\mu_{TCP}.$}
\eea
These formulae are obtained by expanding the equation of state in a
power series of
the expectation value and by taking the coefficient of the linear and cubic
terms, which correspond to the quadratic and quartic terms of the
effective potential. In the formulae above $-z=e^{\frac{\mu_q}{T_c}}$ 
(the fugacity),
$c=2e^{1-\gamma_E}$, $N_c=3$, and $M_0$ is the renormalisation scale.
                  
The curvature of the line of second order phase transitions at
$\mu_q=0$ is $\frac{T_c}{2}\frac{d^2 T_c}{d\mu^2_q}\Big|_{\mu_q=0}=0.101$,
which is somewhat higher than what is obtained on the lattice. (The masses
are different, strictly speaking there is no much sense in comparing). This
can be seen also in the r.h.s. of Fig.~\ref{Fig:mu-T} taken 
from \cite{deForcrand2} in which we
inserted our phase transition line (the green line which ends in the
lower right corner of the figure). The reason for this is that the large
value of the Yukawa coupling lowers the critical
temperature at vanishing chemical potential and bends too strongly the
transition line. We have checked that dynamical generation of the fermion
mass, e.g. solving a gap equation for it, lowers the effective Yukawa 
coupling. This result and the fact that according to Fig.~\ref{Fig:mu-T} the
line of the reduced temperatures lies within 10\% of the lattice result, 
mean that it might be worth thinking about a more complete inclusion of 
fermions, e.g. beyond one loop order. 

By looking at different effective approaches published in the
literature (see a list in \cite{Stephanov}) we can see
that all give low values for the temperature and high values for the
chemical potential at TCP or CEP when compared with the result
of Fodor and Katz \cite{Fodor_Katz}.
% at least one of the values is far away from the lattice result. 
This may be partly due to the fact that 
these results are for two flavors and not 2+1 ( 
it was  emphasised in Ref.~\cite{Philipsen} how sensitive is the
dependence of the
critical value of $\mu_B$ on the value of the quark masses).
But this also raises a conceptual problem of whether 
the QCD phase transition is driven by the chiral symmetry.

As was shown recently in \cite{Braun} using renormalisation group techniques
the inclusion of gluon degrees of freedom proves to be important at
$T\approx 150$ MeV if one wants to compare the equation of state of the
effective model with the one obtained on the lattice.

The conclusion of the resonance gas model (which includes 1026 resonances)
is that the phase transition is driven by the higher excited hadrons rather
than by the light degrees of freedom \cite{RGM}. This model managed to 
reproduce
the temperature dependence of the energy density measured on the lattice. It
is interesting that a pion gas would give only 15\% of the critical energy
density at the transition temperature. Imposing constant energy density the
resonance gas model
also reproduced the phase boundary measured on the lattice using 
Taylor expansion in $\mu_B$.

The discrepancy between the freezout curve and phase boundary obtained with
the resonance gas model shows that there are strong interactions between
baryons and mesons when the density of the baryons exceeds the density of
mesons and without interactions this model is not able to account for how
criticality occurs.

I conclude that the investigation of the phase-diagram in the $\mu_B-T$
plane with an effective 3-flavor model is definitely of interest.

\section{The nature of the soft mode}

In the physical point of the quark mass-plane a critical end point was found
on lattice \cite{Fodor_Katz} at a finite value of the chemical potential and
the question is what causes criticality.  At the critical end point two
minima and a maximum meet which results in a very flat effective potential
of the chiral order parameter $\Phi$. This is shown also by the diverging
peak of the scalar susceptibility $d\Phi/d h$. This flattening of the
effective potential does not necessarily imply the appearance of a gapless
sigma mode (sigma meson with vanishing mass). This is because the
non-interchangeability of two limits
\footnote{Using the relevant finite temperature expressions, one can easily 
prove that although, by definition, the bosonic one-loop bubble integral
with equal masses is related at zero external four-momentum
$q=(\omega,\q)$ to the bosonic tad-pole integral through 
$I_B(q=0,M)=d T_B(M)/d M^2$, this relation holds only
in the $\q$-limit, not in the $\omega$-limit.}
: $\q$-limit 
($\omega\to 0$, then $\q\to \0$), which applies to static
quantities as the curvature of the effective potential, and $\omega$-limit
($\q\to \0$, then $\omega\to 0$), which applies
to the study of a sigma pole at rest on the second Riemann sheet (resonance). 

It was argued in Ref. \cite{Son} that what is important from the point of
view of the signature of the genuinely singular end point is the dynamical
feature rather than the static one. The time during which the correlation
length $\xi$ reaches its equilibrium diverges as $\tau\sim \xi^z$ 
(critical slowing down), where $z$ is the dynamic scaling exponent. 
The finite evolution time limits the correlation
length to $\xi<(\rm{time})^{1/z}$. Because $z>1$ and in heavy-ion
collision the evolution times are of the same order as the spatial size,
it is the finite time of the evolution rather than the spatial size 
limitation ($\xi<$size), which imposes a limit on the value of $\xi$.
This is why, it is important to find the dynamical universality class CEP 
belongs to, according to the classification of \cite{HH_class},
and also the correct identification of the soft mode, the mode whose
characteristic frequency vanishes at CEP, and which determines the 
characteristic time of the evolution.

I refer the reader to the literature in questions which concern 
the hydrodynamical nature of the soft mode, in presence of conserved 
baryon number and energy densities, and on the demonstration that the 
dynamic universality class of CEP is that of model H, i.e. 
the liquid-gas phase transition 
\cite{HH_class}, \cite{Son}, \cite{Fujii_hydro}. Here
I will discuss what can be found about the nature of the
soft mode with model calculations focusing on the scalar order parameter.  

In the O(4) model without explicit symmetry breaking there is a genuine
second order transition accompanied by a change in the symmetry of the
ground state. Here, what is responsible for the critical behaviour and
dynamical scaling is the pole in the sigma channel \cite{Patkos1b}. 
Following the trajectory of the sigma pole (corresponding to the
particle a rest) of the 
analytically continued Green's function on the second Riemann sheet, 
it turns out that the pole arrives on the imaginary axis, splits up
into two poles, and the one that goes to the origin will produce
critical scaling. In the vicinity of the critical point the dynamical
scaling and the equation for the soft mode were given in \cite{Patkos1b} 
(see also the references therein).

In the 2 flavor constituent quark model in the chiral limit, along the line 
of the chiral critical points, the soft mode is also the sigma meson, 
which becomes the degenerate chiral partner of the massless pions. 
The sigma pole on the 2nd Riemann sheet goes to the origin as the order 
parameter $\Phi(T)$ goes to zero. The real and imaginary parts of 
the pole scale as $\textnormal{Re} p_0\sim \Phi(T)$,
$\textnormal{Im} p_0\sim \Phi^2(T)$, where $\Phi\sim(T-T_c)^{\beta}$. 
For $\beta$ the mean-field exponent was obtained \cite{Patkos2}
: $\beta=1/2$ for
the second order line and $\beta=1/4$ at the tricritical point. The case 
of TCP is interesting because it prompts the question of what kind of 
singular behaviour might appear additionally there, where the sigma
meson mode is already gapless (see below).

For the physical pion mass and two flavors no symmetry change of the
ground state accompanies the second order transition at CEP and in 
consequence from the 
point of view of the symmetry the existence of CEP is accidental. 
Both in CQM \cite{Patkos4} and NJL \cite{Fujii} the sigma pole stays massive,
and a peak develops in the scalar isoscalar spectral function in 
the space-like region $\omega<|\q|$. The location of its maximum 
goes to zero as the momentum goes to zero and it diverges in the $\q$-limit.
%the static limit, i.e. first  $\omega \to 0$ and then $|\q|\to 0$.

In the context of the NJL model what is 
responsible for the behaviour seen in the spectral functions is an 
imaginary pole of the scalar susceptibility 
($\chi_{mm}=d^2 F(T,\mu)/d m^2_{ud}=d\langle\bar q q\rangle/d m_{ud}$,
where $F(T,\mu)$ is the free energy density)
that goes to the origin as $|\q|\to 0$ (diffusive mode) \cite{Fujii}. 
In consequence the soft mode is the fluctuation of the scalar density. 
To study the relative weight of the diffusive mode in the susceptibility,
the ratio $R=(\chi_{mm}(0,\0^+)-\chi_{mm}(0^+,\0))/\chi_{mm}(0,\0^+)$
was introduced, where the numerator is the difference between the
$\q$-limit and $\omega$-limit. When the diffusive mode is absent
$R=0$. Because there is no contribution from the diffusive mode in
the second term of the numerator, when the diffusive mode dominates
$R\to 1$. By studying this ratio $R$, it was shown in \cite{Fujii_hydro} 
that the switch of the soft mode from the sigma meson mode to 
the diffusive mode (hydrodynamical mode) occurs at the tricritical point,
where $R=1$. 

%\newpage

\section{The boundary of the first order phase transition in the
$m_\pi-m_K$--plane\label{sec:boundary}}

There are several lattice studies of thermodynamics with varying 
degenerate quarks  $m_u=m_d=m_s$, which show that the value of 
$m_{PS}^c$ on the boundary drops substantially when improved lattice actions 
are used, namely from $m_\pi^c\approx 290$ MeV \cite{Karsch_high_mpc} or
$m_\pi^c\approx 270$ MeV \cite{Christ} to
$m_\pi^c=67(18)$ MeV \cite{Karsch_low_mpc} or ever further down to  
$m_\pi^c< 65$ MeV \cite{MILC_low} (see also the talk by O. Philipsen
at this meeting). 
In view of this result we can think that the linear sigma model
might complement lattice investigations if its coupling parameters can be 
determined accurately because the lighter are the pseudoscalars the
better should become the approximate description based on L$\sigma$M.
This model was claimed to give a low value for the
critical pion mass: $m_\pi^c= 51$  MeV \cite{Meyer-Ortmanns} or
$m_\pi^c= 47$  MeV \cite{Schmidt} (see also \cite{Lenaghan2}).
In these investigations only the Goldstone mass was tuned by varying the
explicit symmetry breaking terms and the other parameters were kept fixed 
at their values determined in the physical point.

We have used ChPT when moving away from the physical point \cite{Patkos3}
for the determination of the pion and kaon decay constants 
and $M_\eta^2=m_\eta^2+m^2_{\eta'}$, to obtain the value of the couplings 
at each point of the $m_\pi-m_K$--plane (see Eqs. (\ref{decayconst_pi}), 
(\ref{decayconst_K}), and (\ref{Meta-extrap})). 
Below I present  
schematically how the process of the parametrisation is performed.

\begin{center}
\bmp[t]{11.0cm}
\begin{tabular}{p{2cm} p{1.0cm} p{0.1cm} p{1.0cm} p{0.2cm} p{0.5cm}}
 & {\bf input:} & & {\bf output:} & & {\bf prediction:} 
\end{tabular}
\begin{equation*}
\begin{aligned}
  \left.
  \begin{aligned}
     \left.
     \begin{aligned}
      f_\pi \, \,  \\f_K  \, \,
     \end{aligned}
      \right\} & \Longrightarrow &  \begin{aligned}
                                    x\quad \, \\
                                    y\quad \,
                                    \end{aligned}
                                    \\
     \left.
     \begin{aligned}
      m_\pi  \,  \\ m_K \,  \\ M_\eta^2   
      \end{aligned}
      \right\} & \Longrightarrow &   \begin{aligned}
                                     g\quad \, \\ 
                                     f_2\quad \\ 
                                     M^2 \, \,\,
                                     \end{aligned}
      \end{aligned} \right\}                             \Longrightarrow  
 \begin{aligned}
m_\eta  \,\,\,\,\,\,\, \quad \quad \quad\\
m_{\eta'}\,\,\,\, \quad \quad \quad \\
\theta_\eta  \quad  \quad \quad \quad \\
m_{a_0}\,\,\, \quad \quad \quad \\
m_{\kappa}\,\, \,\,\quad \quad \quad
\end{aligned} \\
 \begin{aligned}
\left. 
\begin{aligned}
\textnormal{A1 \& ${M}^2$}\,  \\ \textnormal{A2 \& ${M}^2$}\, 
\end{aligned}
\right\}
 & \Longrightarrow & \left.
\begin{aligned}
\mu_0^2\,\,\,\,\, \\ f_1 \, \,\,\,\,
\end{aligned}
\right\}   \Longrightarrow 
\begin{aligned}m_\sigma \\ m_{f_0} \\ \theta_\sigma \end{aligned} \,\,
\qquad\quad\\
\left. 
\begin{aligned}
E_x=0\,\,\, \\ E_y=0 \,\,\,
\end{aligned}
\right\} 
& \Longrightarrow &
\begin{aligned}
h_x \quad\quad \quad\quad\quad \quad\,\,\, \quad\quad \quad\\ 
h_y \quad \quad \quad \quad\quad \quad\,\,\,\quad\quad \quad
\end{aligned}
\end{aligned}
\end{aligned}
\end{equation*}
\emp
\end{center}

The strange and non-strange condensates are determined through the PCAC
relations $x=f_\pi$, $y=(2f_K-f_\pi)/\sqrt{2}$, while the masses
determine the couplings (for the explicit formulae see \cite{Lenaghan1}
or \cite{Patkos3}). The predicted
values of $m_\eta$ and $m_\eta'$ are in remarkable agreement with
the values of these masses given by ChPT up to $m_K=800$ MeV, 
even for vanishing pion mass.
Unfortunately, at tree-level, in the pseudoscalar sector one of the 
quartic couplings, $f_1$ and the mass parameter $\mu_0^2$ appear only in 
the combination $M^2=-\mu_0^2+4f_1(x^2+y^2)$. To disentangle them we have 
to use also the scalar sector. Since the dependence of scalar masses on 
$m_\pi$ and $m_K$ is not known, we need to make assumptions, which
diminish the predictive power of our solution.
This can be seen in Fig.~\ref{fig:PD_A1_A2} which shows the boundary
of the first order transition region on the  $(m_\pi-m_K)$-plane
for various assumptions made for the continuation of the scalar mass spectrum.
We expect that the exact phase boundary curve lies in the light-grey shaded
region.

\begin{figure}[htbp]
\centering{\includegraphics[width=0.6\textwidth]{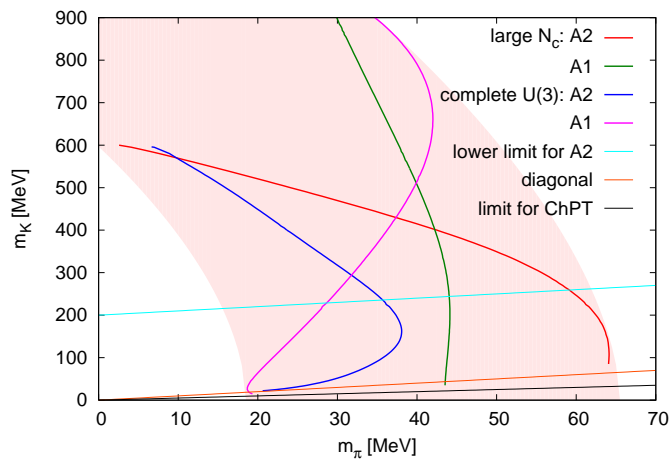}}
\caption{Phase boundary for different alternatives. 
A1: it is assumed that there is no mixing in the scalar $x-y$ sector
($m^2_{\sigma_{xy}}=0$)
A2: it is assumed that the $SU_L(3)\times SU_R(3)$  Gell-Mann--Okubo 
mass formula for the scalars is satisfied in the full $(m_\pi-m_K)$-plane
with the same accuracy as at the physical point.
We consider complete U(3) ChPT and the leading order large-$N_c$ ChPT 
at $\mathcal{O}(p^2)$. Crossover (first order transition) takes places
at the right (left) of the shaded region.}
\label{fig:PD_A1_A2}
\end{figure}

Our rough estimate for the critical pion mass on the diagonal is
$m_\textrm{crit}(\textrm{diag})=40\pm 20$ MeV, in nice agreement with
the results of the latest effective model and lattice studies. In order to be
able to use the tree-level parametrisation we used a quasi-particle
approximation, that is we neglected the finite parts of the
$T=0$ quantum fluctuations. We note, that we were not able to locate the 
TCP on the $m_{u,d}=0$ axis. Recently a work performed in the NJL model 
found the TCP at the critical value of $m_s^c=10$ MeV \cite{Ravagli}, 
which is much lower than expected when extrapolating lattice results.

\section{Conclusions}

In this contribution we reviewed some of the results obtained with effective
models of QCD on the chiral symmetry restoration. We have shown that one can
parameterise the linear sigma model in the entire $m_\pi-m_K$--plane 
employing the results of the chiral perturbation theory. We have
determined the boundary of the first order phase transition line which is
below $m_\pi^c=65$ MeV. For a more reliable determination of the boundary,
information on the quark-mass dependence of the scalar sector is needed.

We have shown that effective models correctly describe the changes of the
meson properties during the chiral transition and indicate that $U_A(1)$ has
to be restored as a prerequisite for restoration of $SU(3)$ chiral symmetry.
The sigma meson is the soft mode at the critical point of the O(4) model
with massless pions, and in both the 2 flavor constituent quark model and
NJL model in the chiral limit along the line of the chiral critical points,
but stays massive at CEP. At CEP the role of the chiral symmetry is rather
secondary, here the soft mode has a hydrodynamical nature.

An analytical determination of the $\mu-T$ phase boundary was given for two
flavors in the chiral limit. The obtained phase diagram is qualitatively
correct, but the transition temperature is low and the value of the chemical
potential at TCP/CEP is apparently too high. Light mesonic degrees of
freedom, usually included into effective theories of QCD do not seem
to be the single driving force behind the transition. In view of this, 
the extension of the 3-flavor chiral quark model to finite density is of 
imminent importance. It would be also interesting to develop 
large N techniques for $SU_L(N)\times SU_R(N)$ 
chiral models, since this approach has some advantage when used together 
with other resummation techniques (2PI or Dyson-Schwinger).

\acknowledgments{
\noindent Work supported by Hungarian Scientific Research Fund (OTKA) under
contract number T046129. Support by the Hungarian Research and Technological 
Innovation Fund, and the Croatian Ministry of Science, Education and
Sports is gratefully acknowledged. The author is supported by OTKA 
Postdoctoral Fellowship (grant no. PD 050015). He would like to thank
A. Patk{\'o}s and P. Sz{\'e}pfalusy for careful reading of the manuscript, 
for useful comments and discussions.}

}
\end{document}